\documentclass{elsart}
\usepackage{amsfonts}
\usepackage{amsmath}

\input{tcilatex}

\begin{document}

\begin{frontmatter}

\title{Generalized Gibbs canonical ensemble:\\A possible physical scenario}

\author{L. Velazquez}

\address{Departamento de Física, Universidad de Pinar del Río \\  Martí 270, Esq. 27 de Noviembre, Pinar del Río, Cuba}

\begin{abstract}
After reviewing some fundamental results derived from the introduction of
the generalized Gibbs canonical ensemble, $\omega \left( E\right)
=\exp \left[ -\eta \Theta \left( E\right) \right] /Z\left( \eta \right) $,
such as the called \textit{thermodynamic uncertainty relation} $\left\langle
\delta \beta \delta E\right\rangle =1+\left( \partial ^{2}S/\partial
E^{2}\right) \left\langle \delta E^{2}\right\rangle $, it is described a
physical scenario where such a generalized ensemble naturally appears as a
consequence of a modification of the energetic interchange mechanism between
the interest system and its surrounding, which could be relevant within the
framework of long-range interacting systems.
\end{abstract}


\end{frontmatter}

\section*{Introduction}

Conventional Thermodynamics and the standard Statistical Mechanics are
inadequate to deal with the called \textit{nonextensive systems}. Loosely
speaking, the non extensive systems are all those systems non necessarily
composed by a huge number of constituents, that is, they could be small or
mesoscopic, where the underlying interactions have a long-range comparable
with or larger than the characteristic linear dimension of the system,
leading in this way to the existence of long-range correlations which do not
support the statistical independence or separability of the large
short-range interacting (\textit{extensive}) systems. The available and
increasingly experimental evidences on anomalies presented in the dynamical
and macroscopic behavior in plasma and turbulent fluids \cite%
{bog,bec,sol,shl}, astrophysical systems \cite{lyn,pie,pos,kon,tor}, nuclear
and atomic clusters \cite{gros1,dago,ato}, granular matter \cite{kud}
glasses \cite{par,stil} and complex systems \cite{sta1} constitute a real
motivation for the generalization of these fundamental physical theories.

The present paper addresses a simple generalization scheme of the
Statistical Mechanics for the Hamiltonian systems which starts from the
Boltzmann-Gibbs Statistics and the consideration of geometric features of
the microcanonical ensemble. As already shown in previous papers \cite%
{vel1,vel2,vel3,toral,vel.geo,vel.flu}, such ingredients allow us a simple
generalization of the formalism of the conventional\ Thermodynamics which is
able to overcome all those difficulties related to the \textit{ensemble
inequivalence phenomenon} or the presence of \textit{negative heat capacities%
}. This results have found an immediately application in the enhancing of
the available Monte Carlo methods based on the canonical ensemble in order
to deal with the existence of the first-order phase transitions \cite%
{vel-mmc,vel.sw}. The main interest of this work is to describe a possible
application scenario of this generalized framework in the context of the
open Hamiltonian systems.

The paper is organized as follows. Firstly, I shall consider in the next
section a brief review of the present generalization scheme and some of
their fundamental results, such as the concept of reparametrization
invariance, the generalized Gibbs canonical ensemble, the \textit{%
thermodynamic uncertainty relation} and its application for the enhancing of
the available Monte Carlo methods based on the standard Statistical
Mechanics. A possible physical scenario for the nature appearance of the
generalized Gibbs canonical ensemble is analyzed in the section \ref%
{relevance}. Finally, some conclusions are given in the section \ref%
{conclusion}.

\section{A review of fundamental concepts and results\label{review}}

Let us limited in the sake of simplicity to the special case of classical
Hamiltonian systems whose thermodynamic description can be performed
starting from microcanonical basis by considering only the total energy $%
E=H_{N}\left( X\right) $ (defined on every point $X$ of the N-body phase
space $\Gamma $). Let $\Theta _{N}\left( X\right) =\Theta \left[ H_{N}\left(
X\right) \right] $ be a bijective and differentiable function of the
Hamiltonian $H_{N}\left( X\right) $. It is said that the function $\Theta
=\Theta \left( E\right) $ is just a reparametrization of the energy. It is
easy to verify that the microcanonical distribution function $\hat{\omega}%
_{M}\left( X\right) $ is \textit{reparametrization invariant }\cite%
{vel1,vel2,vel3}:%
\begin{equation}
\hat{\omega}_{M}\left( X\right) =\frac{1}{\Omega \left( E\right) }\delta %
\left[ E-H_{N}\left( X\right) \right] \equiv \frac{1}{\Omega \left( \Theta
\right) }\delta \left[ \Theta -\Theta _{N}\left( X\right) \right] .
\label{inv}
\end{equation}%
The demonstration of the above identity follows directly from the properties
of the Dirac delta function:

\begin{equation}
\delta \left[ \Theta -\Theta _{N}\left( X\right) \right] =\left( \frac{%
\partial \Theta \left( E\right) }{\partial E}\right) ^{-1}\delta \left[
E-H_{N}\left( X\right) \right] \Rightarrow \Omega \left( \Theta \right)
=\left( \frac{\partial \Theta \left( E\right) }{\partial E}\right)
^{-1}\Omega \left( E\right) ,
\end{equation}%
where $\Omega \left( \Theta \right) =Sp\left\{ \delta \left[ \Theta -\Theta
_{N}\left( X\right) \right] \right\} $ is the microcanonical partition
function in the $\Theta $ energy reparametrization, and $Sp\left[ A\left(
X\right) \right] =\int A\left( X\right) dX$, the phase space integration.

A direct consequence of this symmetry is the reparametrization invariance of
the whole Physics within the microcanonical description, i.e.: the
microcanonical average $\left\langle A\right\rangle =Sp\left\{ \hat{\omega}%
_{M}\left( X\right) A\left( X\right) \right\} $ of any microscopic
observable $A\left( X\right) $ is reparametrization invariant, $\left\langle
A\right\rangle \left( E\right) \equiv \left\langle A\right\rangle \left(
\Theta \right) $. The microcanonical partition function $\Omega $ allows us
the introduction of an invariant measure: 
\begin{equation}
d\mu =\Omega \left( E\right) dE=\Omega \left( \Theta \right) d\Theta ,
\end{equation}%
which leads to an invariant definition of the microcanonical entropy $S=\log
W_{\alpha }$, being $W_{\alpha }=\int_{C_{\alpha }}d\mu $ and $C_{\alpha }$
a subset belonging to a suitable coarsed grained partition $\Im $ of the
phase space $\Gamma $, $\Im =\left\{ C_{\alpha }\subset \Gamma
;\tbigcup\limits_{\alpha }C_{\alpha }=\Gamma \right\} $, i.e.: $C_{\alpha
}\equiv \left\{ X\in \Gamma :~E_{\alpha }-\frac{1}{2}\delta \varepsilon \leq
H_{N}\left( X\right) <E_{\alpha }-\frac{1}{2}\delta \varepsilon \right\} $.
The coarsed grained nature of the microcanonical entropy can be dismissed
whenever the interest system is large enough ($\delta \varepsilon <<E$), so
that, such a thermodynamic potential could be considered as a continuous
scalar function $S\left( E\right) =S\left( \Theta \right) $. The above
reasonings allow us to claim that the microcanonical description can be
performed by using any reparametrization of the total energy since this
geometrical feature does not involve a modification of the underlying
Physics within this ensemble.

The convex or concave character of any scalar function depends on the
specific reparametrization use for describe it. A trivial example is the
concave function $s\left( x\right) =\sqrt{x}$ with $x>0$, which turns a
convex function $s\left( y\right) =y^{2}$ under the reparametrization change 
$x=y^{4}$ with $y>0$. This remarkable observation allows us to claim that
the convex or concave character of the entropy is not a property
microcanonical relevant since the concavity can be arbitrarily modified by
the consideration of energy reparametrizations.

However, the convex character of the microcanonical entropy in terms of the
total energy has taken always as an indicator of the ensemble inequivalence
within the Gibbs canonical ensemble:%
\begin{equation}
\hat{\omega}_{C}\left( X\right) =\frac{1}{Z\left( \beta \right) }\exp \left[
-\beta H_{N}\left( X\right) \right] ,  \label{can}
\end{equation}%
which obviously does not obey the reparametrization invariance of the
microcanonical ensemble (\ref{inv}). The discrepancy between them is
explained by the different underlying physical conditions: microcanonical
description corresponds to an isolated system, while the canonical one
corresponds to an open system. 

Inspired on the reparametrization invariance of the microcanonical
description, some consequences of \textit{assuming the energy
reparametrizations within the framework of the Boltzmann-Gibbs statistics}
were analyzed in previous papers \cite{vel2,vel3,toral,vel.geo}, as example,
the following generalization of the Gibbs canonical ensemble: 
\begin{equation}
\hat{\omega}_{GC}\left( X\right) =\frac{1}{Z\left( \eta \right) }\exp \left[
-\eta \Theta _{N}\left( X\right) \right] .  \label{gce}
\end{equation}

A simple but remarkable consequence of the reparametrization invariance of
the microcanonical description is the reparametrization invariance of the
formal structure of the thermodynamic formalism \cite{vel.geo}, which could
be easily shown throughout the relation between the canonical partition
function $Z$ and the microcanonical partition function $\Omega $: 
\begin{equation}
Z\left( \eta \right) =\int \exp \left[ -\eta \Theta \left( E\right) \right]
\Omega \left( E\right) dE\equiv \int \exp \left[ -\eta \Theta \right] \Omega
\left( \Theta \right) d\Theta ,
\end{equation}%
which leads when the interest system is large enough to the following
Legendre transformation in terms of the energy reparametrization $\Theta $: 
\begin{equation}
P\left( \eta \right) \simeq \inf_{\Theta }\left\{ \eta \Theta -S\left(
\Theta \right) \right\} ,
\end{equation}%
being $P\left( \eta \right) =-\log Z\left( \eta \right) $ the Planck
thermodynamic potential associated to the generalized Gibbs canonical
ensemble (\ref{gce}). This result shows that many results of the Gibbs
canonical ensemble (\ref{can}) can easily extended to the generalized
ensemble with a simple change of reparametrization $\left( E,\beta \right)
\rightarrow \left( \Theta ,\eta \right) $. Particularly, the ensemble
inequivalence is now associated to the existence of convex regions of the
microcanonical entropy $S$ in terms of the energy reparametrization $\Theta $%
, $\partial ^{2}S\left( \Theta \right) /\partial \Theta ^{2}<0$. Obviously,
this result clearly shows us that those inaccessible regions within the
canonical description (\ref{can}) become accessible within the generalized
canonical ensemble (\ref{gce}) by using a suitable energy reparametrization $%
\Theta \left( E\right) $. An direct application of this idea is just the
generalization of the Metropolis importance sample algorithm \cite{metro} in
order to deal with the ensemble inequivalence (see below).

A preliminary vision about the underlying physical conditions leading to the
generalized canonical ensemble (\ref{gce}) is found in terms of the
Information Theory \cite{jaynes}. The generalized ensemble follows from the
use of the well-known Shannon-Boltzmann-Gibbs extensive entropy:%
\begin{equation}
S_{e}\left[ p\right] =-\sum_{k}p_{k}\log p_{k},
\end{equation}%
and the imposition of the constrain:%
\begin{equation}
\left\langle \Theta \right\rangle =\sum_{k}\Theta \left( E_{k}\right) p_{k},
\label{constrain}
\end{equation}%
a viewpoint recently developed by Toral in the ref.\cite{toral}. The analogy
with the Gibbs canonical ensemble clearly suggest us that the generalized
canonical ensemble (\ref{gce}) can be associated to the influence of certain 
\textit{external apparatus} performing a control process on the interest
system by keeping fixed the average (\ref{constrain}) instead of the usual
energy average with $\Theta \left( E\right) =E$. In this case, the external
control apparatus is just a more sophisticate version of the Gibbs
thermostat, where the parameter $\eta $ plays a role completely analogue to
the inverse temperature $\beta $.

Despite of the mathematical analogies, the physical relevance of the energy
reparametrization $\Theta \left( E\right) $\ and the generalized canonical
parameter $\eta $ of the external control apparatus in general is unclear. A
suitable way to understand their meaning is unexpectedly obtained when the
generalized ensemble (\ref{gce}) is used during the Metropolis simulation 
\cite{metro} of the interest system: the using of the generalized canonical
weight $\exp \left[ -\eta \Delta \Theta \right] $ \cite{vel-mmc} instead of
the usual $\exp \left[ -\beta \Delta E\right] $, where $\Delta \Theta
=\Theta \left( E+\Delta E\right) -\Theta \left( E\right) $. Since $\Delta
E\ll E$ when the interest system is large enough, $\Delta \Theta \simeq
\partial \Theta \left( E\right) /\partial E\ast \Delta E$, so that, the
acceptance probability of a Metropolis move can be given by $p\left( \left.
E\right\vert E+\Delta E\right) \simeq \min \left\{ 1,\exp \left[ -\tilde{%
\beta}\Delta E\right] \right\} $,$~$where: 
\begin{equation}
\tilde{\beta}=\eta \frac{\partial \Theta \left( E\right) }{\partial E}.
\label{effect_b}
\end{equation}%
This result clearly indicates that in terms of the energetic interchange
with the interest system, \textit{the external control apparatus} (or
generalized thermostat with parameter $\eta $)\textit{\ is almost equivalent
to a Gibbs thermostat with an effective fluctuating inverse temperature }$%
\tilde{\beta}$. Obviously, such an external control apparatus possesses a
more active role than the usual Gibbs thermostat. It can be shown that the
usual equilibrium condition between the inverse temperature $\hat{\beta}$ of
the thermostat and the corresponding inverse temperature $\hat{\beta}$ of
the interest system takes place only in average, $\left\langle \tilde{\beta}%
\right\rangle =\left\langle \hat{\beta}\right\rangle $, being:%
\begin{equation}
\hat{\beta}=\frac{\partial S\left( E\right) }{\partial E},
\end{equation}%
the microcanonical inverse temperature of the interest system \cite{vel.flu}.

A very important result is obtained by analyzing the correlation function
between the effective inverse temperature $\tilde{\beta}$ of the generalized
thermostat and the total energy $E$\ of the interest system. The dispersion $%
\delta \beta =\hat{\beta}-\left\langle \hat{\beta}\right\rangle $ can be
estimated when the interest system is large enough by using the equation (%
\ref{effect_b}) as follows: 
\begin{equation}
\delta \beta =\eta \frac{\partial ^{2}\Theta \left( \bar{E}\right) }{%
\partial E^{2}}\delta E\Rightarrow \left\langle \delta \beta \delta
E\right\rangle =\eta \frac{\partial ^{2}\Theta \left( \bar{E}\right) }{%
\partial E^{2}}\left\langle \delta E^{2}\right\rangle ,
\end{equation}%
where $\bar{E}=\left\langle E\right\rangle $, and $\delta E=E-\left\langle
E\right\rangle $ is the energy dispersion of the interest system, which is
related with the dispersion $\delta \Theta =\Theta -\left\langle \Theta
\right\rangle $ as follows:%
\begin{equation}
\left\langle \delta \Theta ^{2}\right\rangle =\left( \frac{\partial \Theta
\left( \bar{E}\right) }{\partial E}\right) ^{2}\left\langle \delta
E^{2}\right\rangle .
\end{equation}%
The reparametrization invariance allows us to derive the parameter $\eta $
and the dispersion $\left\langle \delta \Theta ^{2}\right\rangle $ from the
microcanonical entropy in the reparametrization $\Theta $ as follows:%
\begin{equation}
\eta =\frac{\partial S\left( \bar{\Theta}\right) }{\partial \Theta }%
,~\left\langle \delta \Theta ^{2}\right\rangle =-\left( \frac{\partial
^{2}S\left( \bar{\Theta}\right) }{\partial \Theta ^{2}}\right) ^{-1},
\end{equation}%
where $\bar{\Theta}=\left\langle \Theta \right\rangle =\Theta \left( \bar{E}%
\right) $, and therefore: 
\begin{equation}
-\left( \frac{\partial \Theta }{\partial E}\right) ^{2}\frac{\partial ^{2}S}{%
\partial \Theta ^{2}}\left\langle \delta E^{2}\right\rangle =1\text{ and }%
\left\langle \delta \beta \delta E\right\rangle =\frac{\partial ^{2}\Theta }{%
\partial E^{2}}\frac{\partial S}{\partial \Theta }\left\langle \delta
E^{2}\right\rangle .  \label{previous}
\end{equation}%
The using of the transformation rule of the entropy Hessian: 
\begin{equation}
\frac{\partial ^{2}S}{\partial E^{2}}=\left( \frac{\partial \Theta }{%
\partial E}\right) ^{2}\frac{\partial ^{2}S}{\partial \Theta ^{2}}+\frac{%
\partial ^{2}\Theta }{\partial E^{2}}\frac{\partial S}{\partial \Theta },
\end{equation}%
allows us to condense the results expressed in the equation (\ref{previous})
in a final remarkable form: 
\begin{equation}
\left\langle \delta \beta \delta E\right\rangle =1+\frac{\partial ^{2}S}{%
\partial E^{2}}\left\langle \delta E^{2}\right\rangle .  \label{tur}
\end{equation}

The thermodynamic relation (\ref{tur}) constitutes by itself a fundamental
result which generalizes the \textit{fluctuation theory} of the standard
Statistical Mechanics. Generally speaking, the Thermodynamics theory deals
with equilibrium situations where the\ total energy $E$ of the interest
system (microcanonical description) or the inverse temperature $\tilde{\beta}
$ of the thermostat (canonical description) is fixed. However, the equation (%
\ref{tur}) clearly shows that the generalized canonical ensemble (\ref{gce})
is just a natural theoretical framework for consider physical situations
where both $\tilde{\beta}$ and $E$ fluctuate around their equilibrium values.

Taking into account the microcanonical extension of the heat capacity~$C$:%
\begin{equation}
C=\frac{dE}{dT}=-\left( \frac{\partial S}{\partial E}\right) ^{2}\left( 
\frac{\partial ^{2}S}{\partial E^{2}}\right) ^{-1},
\end{equation}%
the reader can understand that the relation (\ref{tur}) constitutes a
suitable generalization of the well-known relation $C=\beta ^{2}\left\langle
\delta E^{2}\right\rangle $ of the conventional Thermodynamics, which allows
us to consider all those microcanonical states with a negative heat capacity 
$C<0$ associated to the convex character of the microcanonical entropy.
Ordinarily, such anomalous macrostates are hidden by the ensemble
inequivalence, which takes place as a consequence of the \textit{%
incompatibility} of the ordinary restriction $\delta \beta =0$ within the
convex regions of the microcanonical entropy, $\partial ^{2}S/\partial
E^{2}>0$. The thermodynamic relation in these cases leads to the following
inequality:%
\begin{equation}
\left\langle \delta \beta \delta E\right\rangle >1,
\end{equation}%
which allows us to claim two remarkable conclusions: (1) \textit{The
anomalous macrostates with a negative heat capacity can be only accessed by
using a generalized thermostat with a fluctuating inverse temperature}; (2) 
\textit{The total energy of the interest system }$E$\textit{\ and the
inverse temperature of the thermostat }$\tilde{\beta}$\textit{\ behave as
complementary thermodynamical quantities within the regions of ensemble
inequivalence}: the energy fluctuations $\delta E$ can not be reduced there
without an increasing the fluctuations of the inverse temperature $\delta
\beta $ of the generalized thermostat, and vice versa. This is the reason
why the result expressed in the equation (\ref{tur}) can be referred as 
\textit{thermodynamic uncertainty relation} \cite{vel.geo}, since it imposes
obviously certain restriction to the determination of the microcanonical
caloric curve $\beta $ \textit{versus} $E$.

The reader can appreciated that the thermodynamic uncertainty relation (\ref%
{tur}) does not make any reference to the energy reparametrizations $\Theta
\left( E\right) $, that is, there is no reference to the underlying
geometric formalism where this result was derived from. Such a remarkable
feature suggests us the general applicability of the thermodynamic relation (%
\ref{tur})\ to the thermodynamic equilibrium between the interest system
under the influence of thermostat. This far-reaching conclusion has a
paramount importance in the generalization of all the available Monte Carlo
methods based on the canonical ensemble in order to overcome all the
difficulties related to the presence of the first-order phase transitions 
\cite{vel.sw}. Apparently, the using of a generalized thermostat with a
fluctuating inverse temperature can be easily combined with any Monte Carlo
algorithm based on the canonical ensemble in order to account for all those
thermodynamic states with a negative heat capacity. \FRAME{ftFU}{4.5455in}{%
2.9983in}{0pt}{\Qcb{Caloric $\protect\beta \left( \protect\varepsilon %
\right) $ and curvature $\protect\kappa \left( \protect\varepsilon \right) $
curves obtained from the Metropolis algorithm and the Swendsen-Wang cluster
algorithm using thermostats associated to the generalized canonical ensemble
(GCE) as well as the canonical ensemble (CE) for the $q=10$ states Potts
model with $L=25$.}}{\Qlb{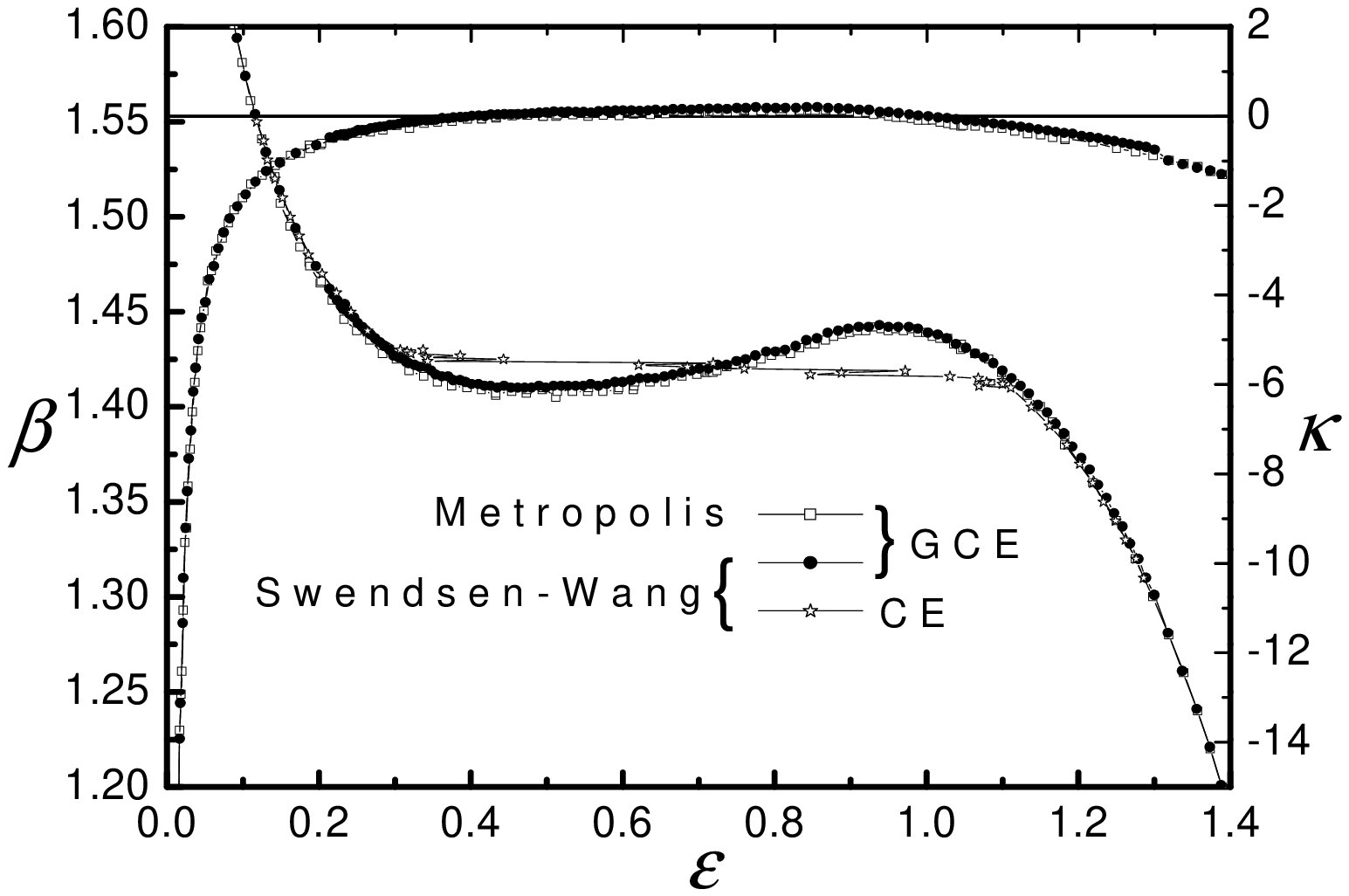}}{compara.eps}{\special{language
"Scientific Word";type "GRAPHIC";maintain-aspect-ratio TRUE;display
"ICON";valid_file "F";width 4.5455in;height 2.9983in;depth
0pt;original-width 6.141in;original-height 4.0369in;cropleft "0";croptop
"1";cropright "1";cropbottom "0";filename 'compara.eps';file-properties
"XNPEU";}}

A numerical evidence suggesting the generality of this idea is shown in the
FIG.\ref{compara.eps}. It shows a comparative study among the caloric curve $%
\beta \left( \varepsilon \right) =\partial s\left( \varepsilon \right)
/\partial \varepsilon $ and the curvature $\kappa \left( \varepsilon \right)
=\partial ^{2}s\left( \varepsilon \right) /\partial \varepsilon ^{2}$ curves
associated to the $q=10$ states Potts model \cite{potts}:%
\begin{equation}
H_{N}=\sum_{\left\langle ij\right\rangle }\left( 1-\delta _{\sigma
_{i},\sigma _{j}}\right) ,
\end{equation}%
on a square lattice $N=L\times L\,$\ with periodic boundary conditions and
only nearest neighbor interactions by using different Monte Carlo methods ($%
\varepsilon =E/N$ and $s=S/N$ are the energy and the entropy per particles
respectively). While the Swendsen-Wang cluster algorithm \cite{wolf} is
unable to account for the anomalous regions with a negative heat capacity by
using the ordinary Gibbs thermostat, the use of a generalized thermostat
with a fluctuating inverse temperature overcomes this difficulty. The reader
can appreciated the remarkable agreement between the results obtained from
the use of the Metropolis importance sample and Swendsen-Wang cluster
algorithm.

\FRAME{ftFU}{4.0413in}{5.2537in}{0pt}{\Qcb{Caloric curve and energy
distributions functions of the $q=10$ states Potts model: Panel a) Phase
coexistence phenomenon and regions of ensemble inequivalence within the
Gibbs canonical description; Panel b) None of these features appear within
the Generalized canonical description. \textexclamdown The first-order phase
transitions are avoidable anomalies within a thermodynamical description
involving energy reparametrizations!}}{\Qlb{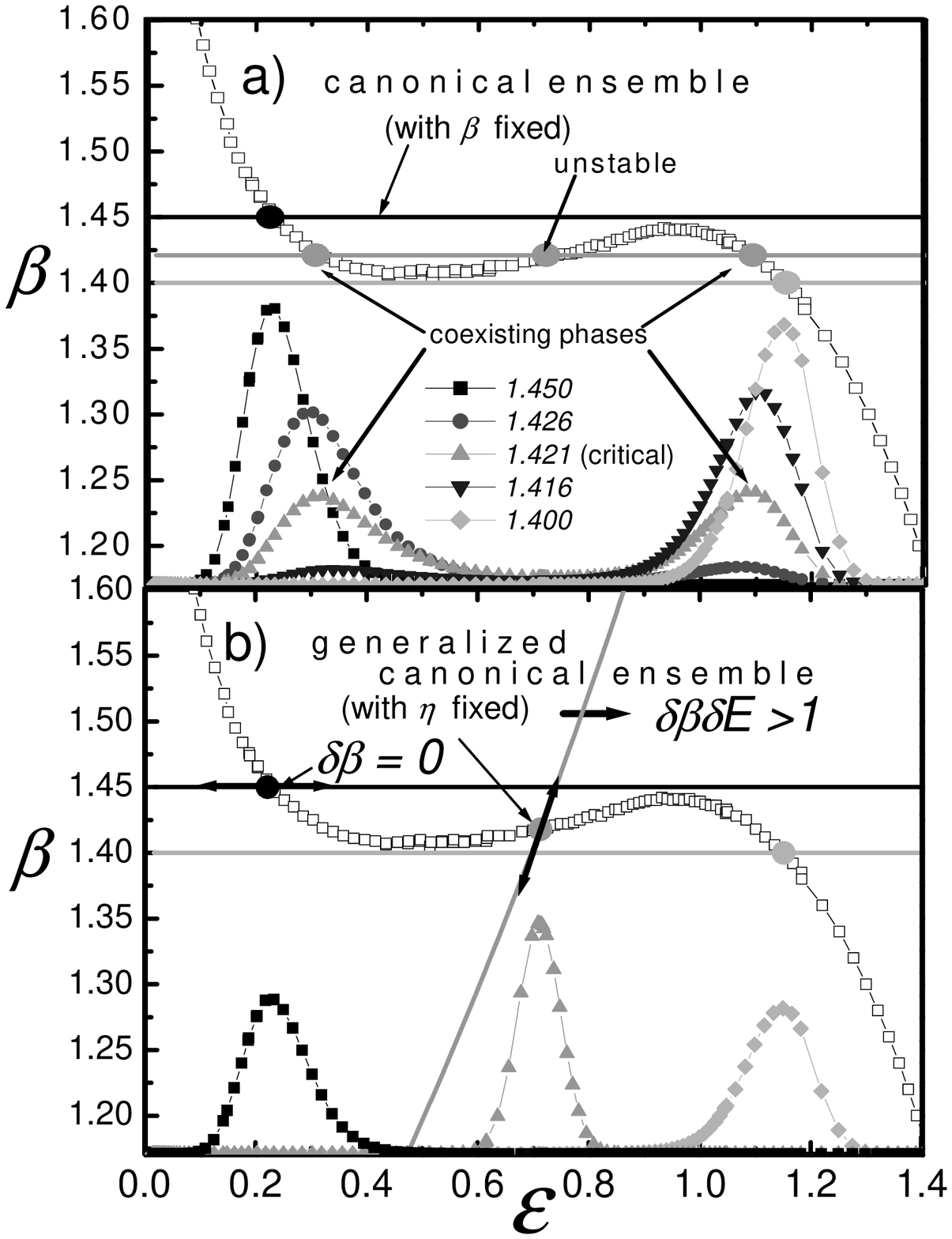}}{comp.thermo.eps%
}{\special{language "Scientific Word";type "GRAPHIC";maintain-aspect-ratio
TRUE;display "ICON";valid_file "F";width 4.0413in;height 5.2537in;depth
0pt;original-width 6.0079in;original-height 7.8196in;cropleft "0";croptop
"1";cropright "1";cropbottom "0";filename 'comp.thermo.eps';file-properties
"XNPEU";}}

The presence of such anomalous regions in the thermodynamical description of
the short-range interacting systems is an indicator of the occurrence of
first-order phase transitions or phase coexistence phenomenon \cite{gros1}.
This feature is clearly illustrated in the FIG.\ref{comp.thermo.eps}.a where
the coexisting phase are reveled as the bimodal character of the energy
distribution function associated to the canonical description in the
neighborhood of the critical temperature. Horizontal lines represent here
the constancy of the inverse temperature of the Gibbs thermostat, $\tilde{%
\beta}=const$. Only those interception points of such horizontal lines with
the microcanonical caloric curve $\beta \left( \varepsilon \right) $
exhibiting a nonnegative heat capacity are accessible within the canonical
description. The constancy of the inverse temperature is replaced in the
generalized canonical ensemble by the constancy of the generalized canonical
parameter $\eta $, which leads to a dependence of the effective inverse
temperature $\tilde{\beta}=\eta \partial \Theta \left( E\right) /\partial E$
of the generalized thermostat on the instantaneous value of the total energy
of the interest system $E$, which is represented in the FIG.\ref%
{comp.thermo.eps}.b. The reader can notice that such a behavior of the
generalized thermostat ensures the interception of only one point of the
microcanonical curve, which eliminates the multimodal character of the
energy distribution functions within this ensemble. Besides, all those
inaccessible macrostates for the canonical ensemble become accessible by
using the present generalized description. This remarkable observation
allows us to claim that \textit{the phase coexistence phenomenon appears as
a consequence of an inefficient control of the external apparatus }%
(thermostat)\textit{\ which is unable to access to all those admissible
microstates of the isolated system }(microcanonical description). Thus, the
first-order phase transitions becomes avoidable thermodynamical anomalies
whenever the energy reparametrizations are involved in the thermodynamical
description, a fact explaining the success of any Monte Carlo algorithms
based on the generalized canonical ensemble.

\section{A possible physical scenario\label{relevance}}

Generally speaking, the generalized canonical description of a given system
is very easy to simulate by using Monte Carlo methods. However, such a
description could be more difficult to implement experimentally since the
usual physical conditions of the environment lead to the relevance of the
Gibbs canonical ensemble (\ref{can}) in the framework of the short-range
interacting systems. 

The previous reasonings do not mean that we can not conceive the
applicability of the generalized canonical ensemble (\ref{gce}) in certain
experimental setup whenever the external "control apparatus" performs a more
active role than the one exhibited by the Gibbs thermostat (while the
thermodynamical state of the Gibbs thermostat is practically unperturbed by
the influence of the interest system under control, the generalized
thermostat changes its effective inverse temperature $\tilde{\beta}$ with
the fluctuations of the total energy $E$ of the interest system). In
principle, there is not \textit{a priory} objection to the designing of
certain experimental arrangement for the implementation of the generalized
canonical ensemble (\ref{gce}).

The way in which the surrounding affect the interest system is crucial for
the relevance of the canonical description (\ref{can}). The well-known Gibbs
argument is based on the separability among the interacting subsystem, which
is supported by the presence of short-range interactions. The energetic
contributions associated to the interactions among the subsystems are
considered here as boundary effects which could be neglected when these
subsystems are large enough. Obviously, \textit{none of the above conditions
can be applicable whenever long-range correlations are involved}.

While the observation of the equilibrium thermodynamical states with a
negative heat capacity in the short-range interacting systems can be only
performed by considering special experimental conditions simulating a
thermostat with a fluctuating inverse temperature, there exist a natural
physical scenario where such an anomaly becomes a very usual behavior: 
\textit{the astrophysical systems}. The existence of equilibrium
thermodynamical states with a negative heat capacity is well-known in the
astrophysical context since the famous Lyndel-Bell work \cite{Lynden}.

The usual way to access to such anomalous macrostates is by using the
microcanonical description, since they are hidden behind of the ensemble
inequivalence of the Gibbs canonical description. The microcanonical
ensemble is the natural thermostatistical description corresponding to the
thermodynamic equilibrium of an isolated system. However, many astrophysical
objects where negative heat capacities are observed do not correspond
necessarily to isolated systems. Contrary, most of these objects can be
considered as open systems driven under the gravitational influence of other
astrophysical objects which are able to interchange energy and particles
with the first one, i.e.: the globular clusters under the influence of the
nearby galaxy \cite{bin}. Contrary to the short-range interacting systems,
the above observation clarifies us that the negative heat capacities are
also relevant for an open astrophysical system under the influence of its
surrounding. Therefore, the framework of the \textit{long-large interacting
systems} could be a natural scenario for the relevance of the generalized
Gibbs canonical ensemble (\ref{gce}).

Despite of the existence of long-range correlations, the generalized
ensemble presupposes certain kind of \textit{separability} of the interest
system from its surrounding. A natural separability appears when the whole
degrees of freedom of the closed system (system + environment) are divided
into internal and collective degrees of freedom. Let us denote the physical
quantities of the interest system and the surrounding by the subindexes $A$
and $B$ respectively. The separability allows us to discompose the total
energy contribution of the closed system $\left( AB\right) $ as follows: 
\begin{equation}
\hat{H}_{AB}\simeq \hat{H}_{A}\left[ a_{AB}\right] +\hat{H}_{B}\left[ a_{AB}%
\right] +\hat{V}_{AB}.
\end{equation}
The terms $\hat{H}_{A}\left[ a_{AB}\right] $ and $\hat{H}_{B}\left[ a_{AB}%
\right] $ represent the energy contribution of the internal degrees of
freedom of the interest system $A$ and the environment $B$ respectively,
where the dependence $\left[ a_{AB}\right] $ represents certain parametric
influence of the collective degrees of freedom on the internal degrees of
freedom. Finally, the term $\hat{V}_{AB}$ consider the pure energy
contribution of the collective degrees of freedom. A simple consequence of
the above assumptions is that the energy interchange between the interest
system $A$ and its surrounding $B$ takes place by means of the influence of
the collective degrees of freedom. While the internal degrees of freedom
obey to a chaotic behavior, I shall assume that the collective degrees of
freedom obey to a quasi-regular dynamics which depends on the energetic
stage of the internal degrees of freedom.

Acoording to the previous hypothesis, the number of microscopic
configurations of the internal degrees of freedom can be estimated by: 
\begin{equation}
\Omega _{A}\left( E_{A};a_{AB}\right) \Omega _{B}\left( E_{B};a_{AB}\right)
dE_{A}dE_{B},
\end{equation}%
where $E_{A}$ and $\Omega _{A}\left( E_{A};a_{AB}\right) $ represents the
internal energy and the density of states of the interest system, as well as 
$E_{B}$ and $\Omega _{B}\left( E_{B};a_{AB}\right) $, the respective
quantities of the surrounding. The total energy of the closed system can be
expressed as follows:

\begin{equation}
E\left( E_{A},E_{B};a_{AB}\right) =E_{A}+E_{B}+\omega \left(
a_{AB};E_{A},E_{B}\right) ,
\end{equation}%
where $\omega \left( a_{AB};E_{A},E_{B}\right) $ represent the energy
contribution of the collective degrees of freedom which drives the energy
interchange between the interest system and its surrounding. Notice that a
dependence of this term on the energetic stage of the internal degrees of
freedom, which appears as a consequence of the influence of the internal
degrees of freedom on the quasi-regular dynamics of the collective degrees
of freedom. Such a term is dismissed in the framework of the short-range
interacting systems because the interactions between the interest system and
the surrounding are just a boundary effect, which is not the case of the
long-range interacting systems. Taken together all the above assumptions,
the number of configurations $W$ of the close system is given by: 
\begin{equation}
W\left( E;a_{AB}\right) =\int \delta \left\{ E-E\left(
E_{A},E_{B};a_{AB}\right) \right\} ~\Omega _{A}\left( E_{A};a_{AB}\right)
\Omega _{B}\left( E_{B};a_{AB}\right) dE_{A}dE_{B},
\end{equation}%
The energy $E_{B}$ can be expressed in terms of the variables $E_{A}$, $E$
and the parameters $a_{AB}$ as follows: 
\begin{equation}
E_{B}=E-E_{A}-\omega \left( a_{AB};E_{A},E_{B}\right) \equiv E-\Theta \left(
E_{A};E,a_{AB}\right) ,
\end{equation}%
where the function $\Theta \left( E_{A};E,a_{AB}\right) $ is just the
solution of the problem: 
\begin{equation}
\Theta =E_{A}+\omega \left( a_{AB};E_{A},E-\Theta \right) .
\end{equation}%
The integration by $dE_{B}$ leads to the following expression:%
\begin{equation}
W\left[ E;a_{AB}\right] =\int \Omega _{A}\left( E_{A};a_{AB}\right) \Omega
_{B}\left[ E-\Theta \left( E_{A};E,a_{AB}\right) ;a_{AB}\right] \nu \left(
E_{A};E,a_{AB}\right) dE_{A},
\end{equation}%
where:%
\begin{equation}
\nu \left( E_{A};E,a_{AB}\right) =1-\frac{\partial \Theta \left(
E_{A};E,a_{AB}\right) }{\partial E}.
\end{equation}%
The approximations $\nu \simeq 1$ and:%
\begin{equation}
\frac{\Omega _{B}\left[ E-\Theta \left( E_{A};E,a_{AB}\right) ;a_{AB}\right] 
}{\Omega _{B}\left[ E;a_{AB}\right] }\simeq \exp \left[ -\eta \Theta \left(
E_{A};E,a_{AB}\right) \right] ,
\end{equation}%
are fully justified by considering the ordinary condition $\Theta \left(
E_{A};E,a_{AB}\right) \ll E$, where $\eta =\partial \log \Omega _{B}\left(
E;a_{AB}\right) /\partial E$ is the microcanonical inverse temperature
associated to the internal degrees of freedom of the surrounding. The
normalization condition of the generalized canonical ensemble:%
\begin{equation}
Z\left( \eta ;a_{AB}\right) =\int \exp \left[ -\eta \Theta \left(
E_{A}\right) \right] \Omega \left( E_{A};a_{AB}\right) dE_{A},
\end{equation}%
is obtained by introducing the reparametrization $\Theta \left( E_{A}\right)
\equiv \Theta \left( E_{A};E,a_{AB}\right) $ and the partition function $%
Z\left( \eta ;a_{AB}\right) =W\left( E;a_{AB}\right) /\Omega _{B}\left(
E;a_{AB}\right) $, which allows us to express the generalized canonical
ensemble as follows: 
\begin{equation}
\hat{\omega}\left( \eta ,a_{AB}\right) =\frac{1}{Z\left( \eta ;a_{AB}\right) 
}\exp \left\{ -\eta \Theta \left( \hat{H}_{A}\left[ a_{AB}\right] \right)
\right\} .
\end{equation}

The present derivation of the generalized canonical ensemble is based on a
separation of internal and collective of degrees of freedom of the closed
system, a procedure that could be relevant within the astrophysical context.
However, the reader can notice that all that is actually needed in order to
support the natural appearance of the generalized ensemble is certain
modification of the mechanism of energy interchange between the interest
system $\left( A\right) $\ and its surrounding $\left( B\right) $. Such a
modification is provided in the present framework by the presence of the
term $\omega \left( a_{AB};E_{A},E_{B}\right) $, which leads to an effective
energy reparametrization\ $\Theta \left( E\right) $ whenever exists a
nonlinear energetic dependence of this function.

The effective inverse temperature of the generalized thermostat associated
to the generalized canonical ensemble:%
\begin{equation}
\tilde{\beta}=\eta \frac{\partial \Theta \left( E_{A}\right) }{\partial E},
\end{equation}%
depends directly on the microcanonical inverse temperature associated to the
internal degrees of freedom on the surrounding $\eta =\partial \log \Omega
_{B}\left( E;a_{AB}\right) /\partial E$, but this quantity is affected by
the incidence of the energetic interchange mechanism leading to the
reparametrization of the internal energy $E_{A}$\ of the interest system.
This feature introduces the following modification of the equilibrium
condition :%
\begin{equation}
\left\langle \tilde{\beta}\right\rangle \equiv \left\langle \frac{\partial
\Theta \left( E_{A}\right) }{\partial E}\right\rangle \frac{\partial \log
\Omega _{B}\left( E;a_{AB}\right) }{\partial E}=\frac{\partial \log \Omega
_{A}\left( E_{A};a_{AB}\right) }{\partial E_{A}}\equiv \left\langle \hat{%
\beta}\right\rangle ,
\end{equation}%
which drops to the usual one whenever the energy contribution $\omega \left(
a_{AB};E_{A},E_{B}\right) $ can be neglected: 
\begin{equation}
\omega \left( a_{AB};E_{A},E_{B}\right) /E_{A}\simeq 0\Rightarrow \partial
\Theta \left( E_{A}\right) /\partial E\simeq 1.
\end{equation}%
As already commented, this is the case of the extensive systems where the
energy contribution characterizing the interaction between the interest
system and its surrounding is just a boundary effect which becomes
negligible in comparison to the internal energy of the interest system. The
energy contribution $\omega \left( a_{AB};E_{A},E_{B}\right) $ could be
comparable to the energy $E_{A}$\ of the interest system in the case of
long-range interacting systems, and therefore, the influence of the
surrounding may become equivalent to a generalized thermostat with a
fluctuating inverse temperature $\tilde{\beta}$.

\section{Conclusions\label{conclusion}}

The geometric framework derived from the reparametrization invariance of the
microcanonical ensemble provides a suitable generalization of the standard
Statistical Mechanics and Thermodynamics which is able to deal with the
phenomenon of ensemble inequivalence and the existence of equilibrium
thermodynamical states with a negative heat capacity \cite%
{vel2,vel3,toral,vel.geo,vel.flu}. Such a development have found an
immediately application to enhance the potentialities of the available Monte
Carlo methods based on the consideration of the canonical ensemble \cite%
{vel-mmc,vel.sw}.

While the usual physical conditions of the environment in the real life
applications of the equilibrium Thermodynamics leads to the applicability of
the Gibbs canonical ensemble (or the Boltzmann-Gibbs distributions), the
practical implementation of the generalized canonical ensemble (\ref{gce})
demands the consideration of a special experimental setup. Nevertheless, a
framework where long-range interactions are involved becomes a possible
physical scenario for the natural appearance of the generalized canonical
description. As already suggested in this paper, the incidence of long-range
correlations could provoke a significant modification of the mechanism for
the energetic interchange between the interest system and its surrounding.
Such a mechanism could be relevant in the astrophysical context, where the
existence of equilibrium states with negative heat capacities is a very
usual phenomenon, in spite of most of astrophysical structures can be
considered as open systems.


\begin{thebibliography}{99}
\bibitem{bog} B. M. Boghosian, Phys. Rev. E \textbf{53} (1996) 4754.

\bibitem{bec} C. Beck, G. S. Lewis, and H. L. Swinney , Phys. Rev. E \textbf{%
63} (2001) 035303(R).

\bibitem{sol} T. H. Solomon, E.R. Weeks, and H. L. Swinney, Phys. Rev. Lett. 
\textbf{71} (1993) 3975.

\bibitem{shl} M. F. Shlesinger, G.M. Zaslavsky, and U. Frisch Eds., \textit{L%
\'{e}vy flights and related topics}, (Springer-Verlag, Berlin, 1995).

\bibitem{lyn} D. Lynden-Bell , Physica A \textbf{263} (1999) 293.

\bibitem{pie} F. Sylos Labini, M. Montuori and L. Pietronero, Phys. Rep. 
\textbf{293} (1998) 61.

\bibitem{pos} L. Milanovic, H. A.Posch and W. Thirring, Phys. Rev. E \textbf{%
57} (1998) 2763.

\bibitem{kon} H. Koyama and T. Konishi, Phys. Lett. A \textbf{279} (2001)
226.

\bibitem{tor} A. Torcini and M. Antoni, Phys. Rev. E \textbf{59} (1999) 2746.

\bibitem{gros1} D.H.E Gross, \textit{Microcanonical thermodynamics: Phase
transitions in Small systems, 66 Lectures Notes in Physics}, (World
scientific, Singapore, 2001) and refs. therein.

\bibitem{dago} M. D'Agostino et al., Phys. Lett. B \textbf{473} (2000) 219.

\bibitem{ato} M. Schmidt et al, Phys. Rev. Lett. \textbf{86} (2001) 1191.

\bibitem{kud} A. Kudrolli and J. Henry, Phys. Rev. E \textbf{62} (2000)
R1489.

\bibitem{par} G. Parisi, Physica A \textbf{280} (2000) 115.

\bibitem{stil} P.G. Benedetti and F.H. Stillinger, Nature \textbf{410}
(2001) 259.

\bibitem{sta1} G.M. Viswanathan , V. Afanasyev , S.V. Buldyrev, E.J. Murphy,
and H.E. Stanley, Nature \textbf{393} (1996) 413.

\bibitem{vel1} L.Velazquez and F.Guzman, \textit{Generalizing the Extensive
Postulates}, e-print(2001) [cond-mat/0107214].

\bibitem{vel2} L.Velazquez and F.Guzman, \textit{The microcanonical theory
and pseudoextensive systems}, e-print(2001) [cond-mat/0107439].

\bibitem{vel3} L.Velazquez and F.Guzman, \textit{Remarks about the
microcanonical description of astrophysical systems}, e-print (2003)
[cond-mat/0303444].

\bibitem{toral} R. Toral, Physica A \textbf{365} (2006) 85.

\bibitem{vel.geo} L. Velazquez and F. Guzmann, \textit{Basis of a non
Riemannian Geometry within the Equilibrium Thermodynamics}, e-print (2006)
[cond-mat/0610712]. Submitted to Phys. Rev. E.

\bibitem{vel.flu} L. Velazquez, \textit{Generalized Fluctuation Theory based
on the reparametrization invariance of the microcanonical ensemble}, e-print
(2006) [cond-mat/0611595]. Submitted to JSTAT.

\bibitem{vel-mmc} L. Velazquez and J. C. Castro Palacio,\ \textit{Metropolis
Monte Carlo algorithm based on reparametrization invariance}, e-print (2006)
[cond-mat/0606727]. Submitted to Phys. Rev. E.

\bibitem{vel.sw} L. Velazquez, \textit{Enhancing Monte Carlo methods by
using a generalized fluctuation theory}, e-print (2006) [cond-mat/0612683].
Submitted to Phys. Rev. E.

\bibitem{metro} N. Metropolis, A. W. Rosenbluth, M. N. Rosenbluth, A. H.
Teller and E. Teller, J. Chem. Phys. \textbf{21} (1953) 1087.

\bibitem{jaynes} E. T. Jaynes, Phys. Rev. \textbf{106} (1957) 620.

\bibitem{potts} J. -S. Wang, R. H. Swendsen and R. Koteck\'{y}, Phys. Rev.
Lett. \textbf{63} (1989) 1009.

\bibitem{wolf} U. Wolff, Phys. Rev. Lett. \textbf{62} (1989) 361.

\bibitem{Lynden} D. Lynden-Bell and R.Wood, MNRAS \textbf{138} (1968) 495;
D. Lynden-Bell, MNRAS \textbf{136} (1967) 101.

\bibitem{bin} J.Binney and S.Tremaine, \textit{Galactic Dynamics} (Princeton
Series in Astrophysics, 1987).
\end{thebibliography}
\end{document}